\documentclass[journal]{IEEEtran}

\usepackage[pdftex]{graphicx}
\ifCLASSINFOpdf
\else
\fi
\usepackage{amsmath}
\usepackage{algorithm, algorithmic}
\usepackage{array}

\usepackage{stfloats}
\usepackage{pgfplots}
\usepackage{tikz}
\usepackage{multirow}
\usepackage{multicol}
\usepackage{amsfonts}
\usepackage{makecell}
\definecolor{ao}{rgb}{0.0, 0.7, 0.0}

\begin{document}
%\renewcommand{\thefigure}{\arabic{figure}}

\title{Reinforcement Learning-based Automatic Diagnosis of Acute Appendicitis in Abdominal CT}

\author{Walid~Abdullah~Al,~\IEEEmembership{Student Member,~IEEE,}
        Il~Dong~Yun,~\IEEEmembership{Member,~IEEE,}
        and~Kyong~Joon~Lee% <-this % stops a space

\thanks{Manuscript submitted on August 30, 2019.
This research was supported by Basic Science Research Program through the National Research Foundation of Korea (NRF), funded by the Ministry of Education, Science, Technology (No. 2017R1A2B4004503).}%
\thanks{W. A. Al and I. D. Yun are with the Department of Computer and Electronic Systems Engineering, Hankuk University of Foreign Studies, Yongin, Gyeonggi, 17035 South Korea (e-mail: walidabdullah@hufs.ac.kr, yun@hufs.ac.kr).}% <-this % stops a space
\thanks{K. J. Lee is with the Department of Radiology, Seoul National University Bundang Hospital, Seongnam, Gyeonggi, 13620 South Korea (email: kjoon@snubh.org).}}

\markboth{}%
{Shell \MakeLowercase{\textit{et al.}}: Bare Demo of IEEEtran.cls for IEEE Transactions on Magnetics Journals}


\maketitle

% for Transactions on Magnetics papers, we must declare the abstract and
% index terms PRIOR to the title within the \IEEEtitleabstractindextext
% IEEEtran command as these need to go into the title area created by
%\IEEEtitleabstractindextext{%
\begin{abstract}
Acute appendicitis characterized by a painful inflammation of the vermiform appendix is one of the most common surgical emergencies. Localizing the appendix is challenging due to its unclear anatomy amidst the complex colon-structure as observed in the conventional CT views, resulting in a time-consuming diagnosis. 
End-to-end learning of a convolutional neural network (CNN) is also not likely to be useful because of the negligible size of the appendix compared with the abdominal CT volume.
With no prior computational approaches to the best of our knowledge, we propose the first computerized automation for acute appendicitis diagnosis. 
In our approach, we utilize a reinforcement learning agent deployed in the lower abdominal region to obtain the appendix location first to reduce the search space for diagnosis. Then, we obtain the classification scores (i.e., the likelihood of acute appendicitis) for the local neighborhood around the localized position, using a CNN trained only on a small appendix patch per volume.
From the spatial representation of the resultant scores, we finally define a region of low-entropy (RLE) to choose the optimal diagnosis score, which helps improve the classification accuracy showing robustness even under high appendix localization error cases. In our experiment with 319 abdominal CT volumes, the proposed RLE-based decision with prior localization showed significant improvement over the standard CNN-based diagnosis approaches. 
  
\end{abstract}



%\IEEEdisplaynontitleabstractindextext

% \fi
\IEEEpeerreviewmaketitle

\section{Introduction}
\IEEEPARstart{A}{cute} appendicitis is identified by a severe inflammation of the vermiform appendix.
This prevalent syndrome among young adults and children falls under the most common surgical emergencies \cite{BHANGU20151278}.
Despite the urgency, physicians face challenges in diagnosing the appendicitis with abdominal CT volumes.
A major challenge is the localization of the appendix as it lacks anatomical clarity in the conventional CT views. Appendix is a tiny projection from the caecum, having a high structural mobility and varied orientation \cite{SELLARS2017432, DESOUZA2015212}. Thus, appendix localization in the convoluted colon becomes a time-consuming process, resulting in a difficult diagnosis. Therefore, a computational diagnosis approach with automated appendix localization can be a useful guide to the physicians to ease the diagnosis process.

% prior works: no prior approach for appendcitis diagnosis

\iffalse
end-to-end diagnosis using full image
 - ex) \cite{kneemri2018, lakhani2017deep}
 - 
ROI-based diagnosis

Prior detection of the ROI
 - patch-based

\fi
\subsection{Related works}
To the best of our knowledge, no prior computational approaches exist for preoperative appendix localization or appendicitis diagnosis. Convolutional neural network (CNN) with its hierarchical and discriminative feature learning has shown promising outcome in image-based diagnosis. It has also enabled several end-to-end diagnosis frameworks directly using the entire image \cite{kneemri2018, lakhani2017deep}. However, such end-to-end learning is not likely to provide satisfactory performance in the context of appendicitis because abdominal CT volumes are relatively huge and can exhaust the CNN computation. Most importantly, the appendix size is negligible even when compared with the lower right abdomen only, therefore, would hardly influence the decision competing with numerous irrelative features throughout the volume. 
\par

The most common approach is to use a small region-of-interest (ROI) to train the network, instead of using the whole image \cite{sahiner2019deep}. Such approaches are prevalently used in the previous works \cite{huynh2016digital, samala2018evolutionary, samala2017multi} and useful to focus only on the target features to train the diagnosis model. However, these approaches generally concentrate on and begin with a given ROI (e.g., masses, tumors, etc.) without detailing the procedure of obtaining the ROI. For diagnosing the appendicitis, the process of detecting the appendix is also important because no other appendix localization approaches exist currently. Moreover, localizing the appendix is usually the major task to the physicians, which slows down the diagnosis, as previously mentioned in the beginning of this section. 
\par
\iffalse
belharbi2017spotting - l3 slice
dou2017multilevel - muti-level context
yap2018automated - 

\fi
Convolutional neural networks have successfully been used to detect a target or suspicious regions useful for diagnosis. The usual approach is to use the patch-based binary classification model, where a CNN is trained only on small patches sampled from the image \cite{belharbi2017spotting, dou2017multilevel, yap2018automated, setionodule}. Image-to-image heatmap regression models are avoided because the suspicious regions are usually sparse. For testing with the patch-based model, a sliding window strategy is usually performed to obtain the candidate patches, which can be costly for a 3D medical image. Recently, converting the trained patch-based CNN model into a fully convolutional network (FCN) during testing showed significant reduction of the cost \cite{doumicrobleeds}. Nevertheless, such patch-based approach can suffer from the \textit{sample selection bias} in case of appendix detection because the positive patches (covering the appendix) would be very few. Moreover, passing the whole abdominal CT into the FCN is also not feasible.     

\begin{figure*}
\centering\includegraphics[scale=1.0]{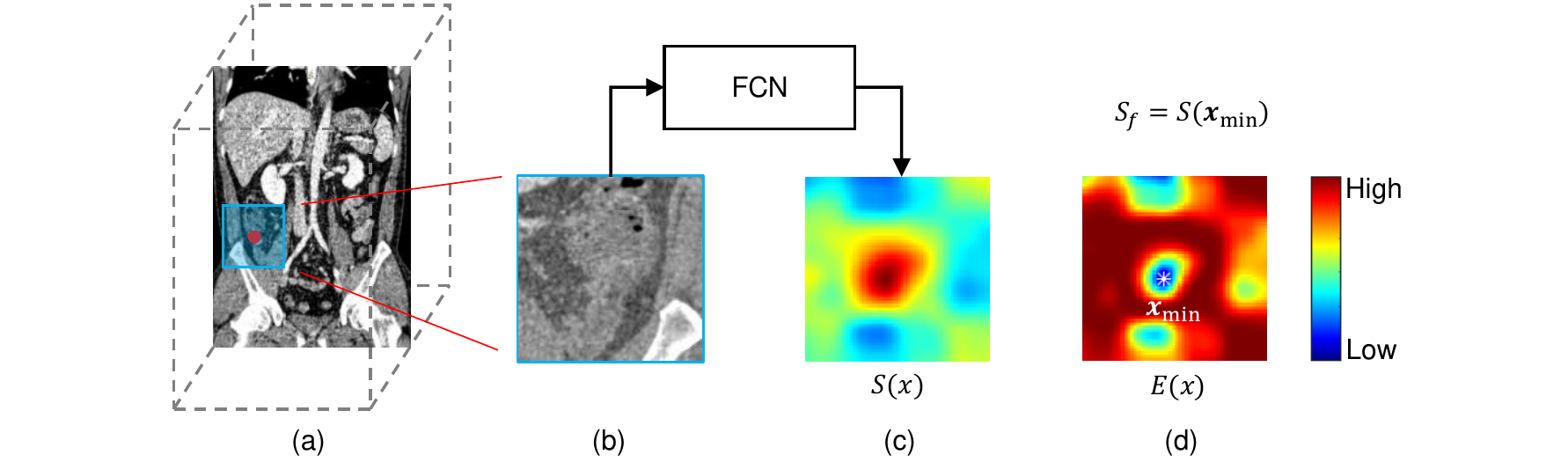}
\caption{2D illustration of the proposed diagnosis framework. (a) Localization of the appendix in the abdominal CT using RL, red dot indicating the localized position. (b) Cropping the neighborhood ROI about the localized position. (c) Score map estimation for the cropped neighborhood. (d) Finding the local minima $\boldsymbol{x}_\text{min}$ of the RLE in the corresponding entropy map, the white asterisk (*) symbol indicating the detected local minima. The final diagnosis score $S_f$ is determined by taking the score at $\boldsymbol{x}_\text{min}$.}
\label{fig:method}
\end{figure*}

\subsection{Proposed approach}  
In this paper, we propose an automatic approach for diagnosing the acute appendicitis, where the appendix location is first inferred by a reinforcement-learning (RL) agent to reduce the search space for diagnosis. For a small neighborhood about the inferred position, classification scores as the probability of the appendicitis are obtained from a CNN classifier, which is trained only on a small appendix patch per volume. Eventually, we propose a region of low-entropy (RLE) in the resultant score map to obtain the final probability of acute appendicitis. Fig.~\ref{fig:method} illustrates the proposed diagnosis framework using a 2D saggital view. Prior localization of the appendix using RL helps avoid exhaustive scanning of the whole CT volume. For a test volume, tracking the optimal appendix patch input for diagnosis using only the localized position is hard. The proposed RLE provides a promising solution for obtaining the score of the optimal appendix patch from the neighborhood of the localized appendix, even in the case of high localization error. Our contribution can be summarized as follows:
\begin{itemize}
\item We propose the first fully automatic diagnosis method for acute appendicitis in abdominal CT.
\item We suggest the RL-based prior localization to overcome the computational and memory cost, and improve the sample selection bias problem, by reducing the candidate search space.
\item We introduce the RLE for robustly identifying the acute appendicitis by pooling the optimal score from the neighborhood of the localized appendix.
\item The proposed RLE also allows for a smaller training patch-size, which has a major contribution to improve the classification accuracy by solely focusing on the appendix. 

\end{itemize}

We organize the rest of the paper as follows. Section~\ref{sec:localization} presents the formulation for the appendix localization agent. In Section~\ref{sec:diagnosis}, we describe the diagnosis of acute appendicitis, defining the RLE from the CNN classifier prediction. We present our experimental evaluation of the proposed diagnosis method in Section~\ref{sec:results}. Finally, we conclude with Section~\ref{sec:conclusion}, presenting our final remarks.

\section{Appendix Localization Using RL}
\label{sec:localization}
\iffalse
localization
    - key to automation
    - reduce search space
    - effective learning with smaller ROI
\fi
The localization of the appendix is the key to the automation of our proposed diagnosis method. The initial estimation of the appendix location helps avoid the costly scanning of the whole abdominal CT and narrows down the search space for diagnosis. It also contributes to learning the diagnosis model (i.e., CNN) effectively. In a large ROI setting for model input, the appendix can hardly influence the learning as compared with other irrelative features that surround it, because of its tiny structure. Prior localization allows reducing the ROI to focus more on the appendix. 
\par

\iffalse
Prior Localization
	- separated appearance learning and search scheme
RL-based localization
	- a navigating agent
\fi

Previous localization works in medical images usually suggest two isolated and sequential steps, where an appearance model is learned first to represent the embedded image context, which is followed by an object search policy \cite{ghesu2019multi}. RL-based approaches have shown remarkable outcome by combining these two steps into a single task for an interactive agent \cite{ghesu2016artificial, ghesu2019multi, alansary2019evaluating}. Whereas the usual approaches perform an exhaustive search, the RL-agent learns an optimal path to navigate to the target in the voxel space. Therefore, localization becomes notably efficient.
\par

\iffalse
Proposal
	- 
\fi

Based on the reformulation in \cite{ghesu2016artificial, ghesu2019multi, alansary2019evaluating}, we utilize an RL agent for localizing the appendix. The agent attempts to update its position sequentially by using its optimal policy for the observed state at the corresponding position, eventually converging to the target location. Usually, the initial position is randomly sampled from the entire volume because no prior information is known. This forces the agent to learn the optimal policy for the full-state space of the vast CT volume. However, we can limit the sample space to the lower-right corner of the abdominal CT because appendix usually lies in the lower-right abdomen. Moreover, we can also reduce the explorable environment for the agent to the lower-right part only, reducing the state-space by 8 times. We define the target location to be the appendiceal base, which is the opening of the appendix from the caecum. Though the centroid of the appendix could be a good target to fit a better ROI around the appendix during the diagnosis, localization becomes difficult because of its mobility. The base, on the other hand, has more consistent and well-distinguishable features. Let us denote this target location by $\boldsymbol{x}_\text{apx}$. Fig.~\ref{fig:rl} illustrates our RL-based localization of the appendix.
\par

\begin{figure}[!t]
\centering
\includegraphics[scale=1.0]{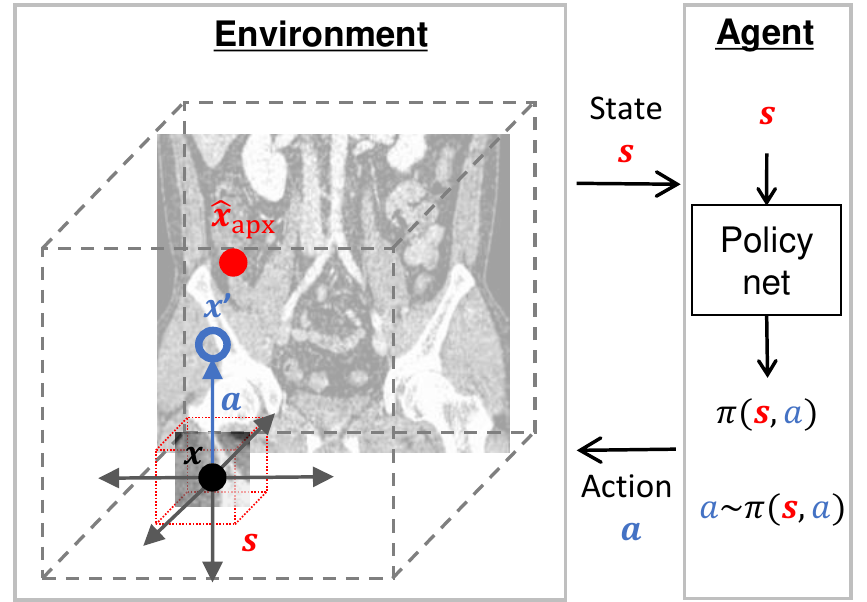}
\caption{Appendix localization using RL. At any position $\boldsymbol{x}$, the agent observes the current state $\boldsymbol{s}$ (i.e., sub-volume centered at $\boldsymbol{x}$) and takes an action $a$ based on its optimal policy $\pi(\boldsymbol{s},a)$ for the current state, thereby moving to a neighboring position. Sequentially making such moves, it eventually converges to the target appendix position.}
\label{fig:rl}
\end{figure}

\iffalse
formulation
	- state
	- 

\fi

Defining action as the unit steps along the Cartesian axes, we allow the agent to choose from six actions to update to a neighboring voxel (considering 6-connectivity). The state for a given position $\boldsymbol{x}$ is defined to be a small sub-volume centered at $\boldsymbol{x}$. The size of the sub-volume is set to $50\times 50\times 50$, which is similar to the previous RL-based localization framework proposed for a variety of anatomical landmarks \cite{ghesu2016artificial, ghesu2019multi, alansary2019evaluating}. The goal of the agent is to optimize its policy so that it can map optimal action for a given state to move towards the target successfully. For training, a positive reward is fed for an action moving the agent closer to the target. Thus, the binary reward signal $R$ can be expressed as:
\begin{equation}
R(\boldsymbol{x},a,\boldsymbol{x'}) = \text{sign} (||\boldsymbol{x-x_\text{apx}}||_2 - ||\boldsymbol{x'-x_\text{apx}}||_2)
\end{equation}
for an agent moving from $\boldsymbol{x}$ to $\boldsymbol{x'}$ by taking action $a$. We adopt the actor-critic approach to learn the optimal behavior effectively, as detailed in our previous work \cite{al2018actor}. Actor-critic learning combines the advantages of both the policy and value-based learning approaches, providing more effective learning. In such frameworks, we learn both the policy and value functions, where policy $\pi(\boldsymbol{s}, a)$ is the optimal action distribution and value $V(\boldsymbol{s})$ is the expected long-term cumulative reward for a given state $\boldsymbol{s}$. Here, $a$ depicts the action.  
\par
The policy and value functions are usually parametrized by using deep neural networks. Policy and value networks can be represented as two individual fully connected stacks after a common CNN. During training, the network parameters are randomly initialized. The agent gathers episodes of transition experiences as tuples $(\boldsymbol{s}, a,\boldsymbol{s'}, r)$ containing original state $\boldsymbol{s}$, chosen action $a$, transitioned state $\boldsymbol{s'}$ and the corresponding reward $r$. Sampling from the gathered experience, it updates the policy and value network parameters. The objective functions and the detailed training procedure is similar to \cite{al2018actor}. Whereas the ultimate goal is to learn the optimal policy $\pi(\boldsymbol{s}, a)$, the value function $V(\boldsymbol{s})$ is only used for reducing the variance during learning. For a test volume, the agent starts at a position $\boldsymbol{x}$ in the lower-right corner and move to a neighboring position by taking action $a$, where $a$ is sampled from the learned policy $\pi(\boldsymbol{s}, a)$ for the state $\boldsymbol{s}$ at position $\boldsymbol{x}$. After repeatedly allowing it to take such moves, we finally take the converged position as the estimated appendiceal base location. 
\par

\section{Acute Appendicitis Diagnosis}
\label{sec:diagnosis}
Following the previous diagnosis approaches, we use 3D CNN to obtain the likelihood of acute appendicitis. Using the inferred appendiceal base location, a possible way to perform the diagnosis is to take an ROI enclosing the appendix. However, this ROI would be large considering the varying localization results caused by the complex anatomy of the appendix. As presented later in Section~\ref{sec:results}, the size of the ROI greatly influences the diagnosis performance. Therefore, to ensure better learning of the CNN model, we take a small ROI (i.e., patch) around the expert-annotated appendiceal base. Thus, we take a single appendix patch from each volume to train our CNN model. We denote the proposed diagnosis model by $D$. 
\par
The RL-localized position for a test volume can be different from the expert-annotated ostium, which can cause problems in finding the desired patch for diagnosis from the inferred position. To tackle this problem, we propose to obtain the acute appendicitis score (from the trained CNN) for the local neighborhood around the localized position. However, the question of combining the neighborhood scores to obtain the final diagnosis remains. Simple voting approaches for ensemble may not result in an accurate diagnosis because the non-appendix patches would be dominating the neighborhood. Training the CNN using multiple neighborhood patches is also not desired because of the possible inclusion of irrelative features. Keeping the single appendix patch per volume criterion for training, we propose to detect the region of low entropy to confidently obtain the final diagnosis from the neighborhood.
\par 
\subsection{Region of low entropy}
We define the RLE in the spatial score-map obtained for the neighboorhood voxels by passing the corresponding patches to the CNN model $D$. Thus, the score $S:\mathbb{X} \rightarrow [0,1]$ can be expressed as: 
\begin{equation}
S(\boldsymbol{x}) = D(\boldsymbol{p_x})
\end{equation}
where $\boldsymbol{p_x}$ denotes the patch centered at $\boldsymbol{x}$, $\mathbb{X}$ indicates the voxel space, and $D$ gives the probability of acute appendicitis for a given patch. The entropy $E:\mathbb{X}\rightarrow \mathbb{R}_+$ for a given probability score can be calculated as follows: 
\begin{equation}
E(\boldsymbol{x}) = - S(\boldsymbol{x}) \log S(\boldsymbol{x}) - (1- S(\boldsymbol{x}))\log(1- S(\boldsymbol{x})) 
\end{equation}
Denoting the inferred appendix location as $\hat{\boldsymbol{x}_\text{apx}}$, we obtain the score map and entropy map for the neighborhood voxels as:
\begin{equation}
\begin{split}
S_\text{map} &= \{S(\boldsymbol{x}) | \boldsymbol{x} \in \text{neihborhood of } \hat{\boldsymbol{x}_\text{apx}}\},\\
E_\text{map} &= \{E(\boldsymbol{x}) | \boldsymbol{x} \in \text{neihborhood of } \hat{\boldsymbol{x}_\text{apx}}\}
\end{split}
\end{equation}
respectively. The corresponding neighborhood patches for computing the score map can be obtained using the sliding window approach. However, this approach of repeatedly cropping 3D patches ignoring the redundant voxels between the adjacent patches has high computational cost. In Appendix~\ref{ap:fcn}, we present an efficient score map estimation approach based on the fully convolutional network (FCN).
\par
Entropy conveys the certainty of a classifier. For example, if a model is trained to classify between cars and trucks, it would give confident class-scores (with bigger margin) for the known car and truck features, resulting in low entropy. On the other hand, the score-margin between the two classes would be low for an input with unknown features (e.g., tree, animal, etc.), resulting in high entropy. Therefore, the entropy near the known object in a scene is lower. The entropy gets higher as we translate away from the object. Thus, a semi-convex low-entropy region can be found with its local minimum near the object that the classifier is trained on. We call this the RLE.
\par

\begin{figure}
\centering\includegraphics[width=0.45\textwidth]{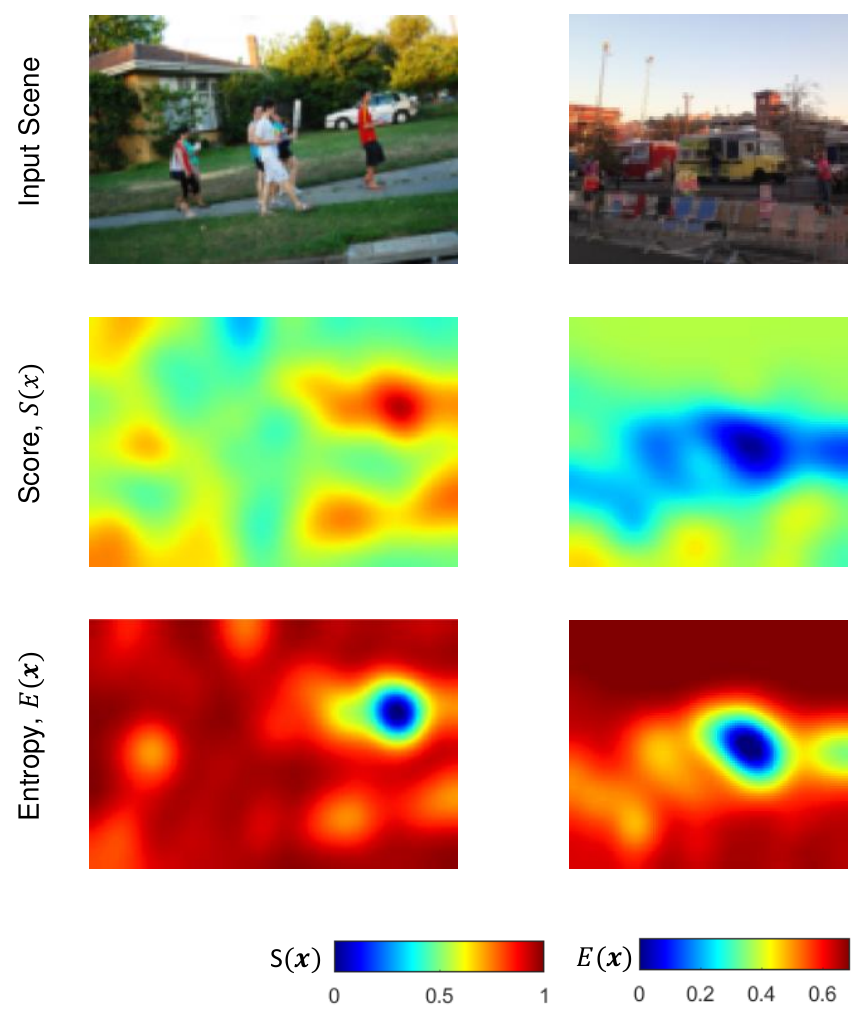}
\caption{RLE visualization in example scenes from PASCAL-VOC and MS-COCO datasets. The presented entropy map is obtained from a classifier trained to classify between car and truck images in CIFAR-10 dataset. For the input scene having car (left), the RLE is observed near the car object in the entropy map, whereas a high entropy is observed elsewhere because of the uncertainty of the classifier. Similar characteristic is observed for the scene having truck (right).}
\label{fig:rle_cifar}
\end{figure}

To render the characteristics of the proposed RLE, we trained a CNN model to classify between car and truck-images in CIFAR-10 dataset \cite{krizhevsky2009learning}. We applied the trained model on two scenes (one having a car object and the other having a truck object), and obtained the corresponding entropy maps. The two scenes are obtained from the PASCAL-VOC \cite{pascalvoc2012} and the MS-COCO dataset \cite{lin2014microsoft}, respectively. The resultant entropy maps are shown in Fig.~\ref{fig:rle_cifar}. The RLE (blue-colored region in the entropy map in Fig.~\ref{fig:rle_cifar}) is prominent near the car and truck compared with the other parts of the scene, where the local minima of the RLE provide the most confident classification.
\par
Because our diagnosis model is trained on the appendix patch centered at the appendiceal base, the RLE is more likely to appear near the appendiceal base. Therefore, we can navigate to the optimal patch for traversing the CNN by finding the RLE, though the RL agent cannot detect the base location exactly for a test volume. To obtain the most certain classification, we select the score at the local minimum of the RLE to be the final diagnosis score, which can be expressed as follows:
\begin{equation}
\begin{split}
S_f &= S(\boldsymbol{x}_\text{min})\\
\boldsymbol{x}_\text{min} &= \arg \min_{\boldsymbol{x}} {E(\boldsymbol{x}), \boldsymbol{x} \in \text{RLE}} 
\end{split}
\end{equation}    
However, multiple RLEs may exist depending on the learnt discriminative appendix feature which may be found in other parts of the complex colon structure. To retrieve the desired RLE at the appendix, we simply weight the local minima of the RLEs by the distance from the localized base position $\hat{\boldsymbol{x_\text{apx}}}$, and choose the one with the minimum weighted entropy.

\section{Results and Discussion}
\label{sec:results}
\iffalse
data
- 
\fi
We evaluated the proposed diagnosis method by performing a 10-fold cross validation on $319$ pre-contrast abdominal CT volumes, which are also used by \cite{kim2012low}. All of the $319$ patients had undergone the CT examination for suspected appendicitis, whereas $116$ volumes were diagnosed to actually have acute appendicitis. The CT volumes had about $476$ axial slices on average, where each axial slice had $512\times 512$ voxels. The average pixel spacing in the axial slice was about $0.60$ mm on both right-to-left and anterior-to-posterior axis. The spacing between slices was $1.0$ mm. The network description of the proposed CNN-classifier for diagnosis is mentioned in Table~A-I. 
\par

\iffalse
localization
dependency/sensitivity on localization
classification
\fi
First, we discuss the localization performance of the RL-agent because this is the basis of the proposed diagnosis method. Then, we present the performance of appendicitis classification from manually annotated ROI around the appendix, where we also show the high sensitivity of the classification to the size of the ROI (i.e., input patch-size). Finally, we present the comparative evaluation of the proposed RLE-based decision of the acute appendicitis from the automatically localized appendix position.

\subsection{Initial Localization of the Appendix}
Performance analysis of the RL-based appendiceal base localization is important because the entire diagnosis method depends on it. We evaluate the performance in terms of the localization error, which is defined as the Euclidean distance of the agent-localized position with respect to the manually annotated appendiceal base location. Initiating near the lower-right corner of a test abdominal CT volume, the agent was allowed to take upto 300 subsequent steps. We obtained the localized appendix position by taking the expectation of the last few positions stepped by the agent. The hyperparameters for training the RL-agent were similar to \cite{al2018actor}. The average localization time was about $50$ milliseconds, as tested on a Geforce GTX Titan Xp GPU, and a 3.60 GHz single-core CPU.

\setcounter{table}{1}

\begin{table*}[b!]
\caption{Comparative Diagnosis Performance of the Proposed Method.}

\centering
\renewcommand{\arraystretch}{1.5}
\setlength{\tabcolsep}{15pt}
\begin{tabular}{l  c c c}
%\multicolumn{4}{|l|}{\bf Heading1} & \multicolumn{4}{|l|}{\bf Heading2}\\ \hline

\hline
{Performance measure} & {\bf RL+FCN+RLE} & { RL+CNN}  & {CNN} \\
\hline
AUC 
& $\mathbf{0.961\pm 0.018}$ 
& $0.698\pm 0.069$ 
& $0.715\pm 0.034$ \\
Sensitivity 
& $\mathbf{0.910 \pm 0.059}$ 
& $0.671 \pm 0.074 $
& $0.669\pm 0.077$\\
Specificity 
& $\mathbf{0.926\pm 0.069}$ 
& $0.647 \pm 0.086 $
& $0.688\pm 0.045$\\

\hline

\end{tabular}

\label{tab:final}
\end{table*}

\setcounter{table}{0}

\begin{table}[t!]
\caption{Localization Error Distribution of the RL Agents for Estimating the Base and Centroid Positions of the Appendix. Localization error is presented as the Euclidean distance of the localized position from the manually-annotated appendix-position.}

\centering
\renewcommand{\arraystretch}{1.5}
\setlength{\tabcolsep}{15pt}
\begin{tabular}{ l c c}
\hline
 \multirow{2}{*}{ \bf Target Position (Case)} & \multicolumn{2}{c}{\bf Localization Error (mm)}\\
& Mean $\pm$ SD & Median\\
\hline
Base (Appendicitis)& $7.31\pm 5.04 $ & $7.16$ \\
Base (Non-Appendicitis)& $ 8.93\pm 5.82 $ & $7.71$ \\
\bf Base (Overall) & {\bf $\mathbf{8.34 \pm 5.04}$} & {\bf $\mathbf{7.42}$}\\
Centroid (Overall) & $15.11 \pm 8.25$ & $12.04$\\
\hline

\end{tabular}

\label{tab:loc}
\end{table}

We have presented the localization error distribution of the proposed RL agent in Table~\ref{tab:loc}. We have also presented the error for the acute appendicitis and non-appendicitis cases separately. The overall error for localizing the appendiceal base was about $8.34 \pm 5.04$ mm with a median of $7.42$ mm. We also trained a separate agent for localizing the centroid of the appendix. We have included the performance of such agent in Table~\ref{tab:loc} to present the difficulty compared to localizing the base as we proposed. Localizing the centroid has an error of $15.11\pm 8.25$ mm, which is significantly larger than the base localization error. This is because appendiceal base has a comparatively more consistent feature than the centroid. Appendiceal base can be identified as the joint of the caecum and appendix, whereas the centroid position can vary with the mobile structure of the appendix and hardly be described as a specific feature-point. Therefore, using the base as the target position, as proposed, is more useful for a robust diagnosis. \par 
%Moreover, we implemented and present the result for standard heatmap regression approach, as described in \cite{payer2016regressing} to localize the base. The RL-based localization showed remarkably better performance compared with such approach. 

\iffalse
- does bigger patch size increase the AUC, 
\fi

\subsection{Classifier-dependency on Patch-size}

Considering the observed localization error, the input patch size for diagnosis should be large enough so that the patch obtained about the localized position can sufficiently accommodate the appendix. However, enlarging the patch size also induces the risk of including irrelative non-appendix features. To analyze the dependency on the patch-size, we tested the performance of the CNN-based diagnosis model assigning different sizes for the input patch centered at the base. We used the manual annotation of the base to obtain both the training and testing patches in order to evaluate the standalone classification performance (without mixing the RL-based localization). For each patch-size, we report the area under the receiver-operating characteristics curve (AUC) as the classifier performance.

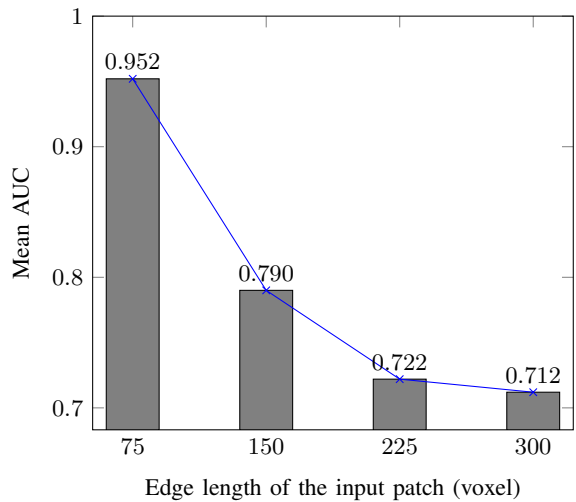
\begin{figure}[]

\pgfplotsset{
width=0.9\columnwidth,
height=0.8\columnwidth,
}
%\centering
\begin{tikzpicture}
\tikzstyle{every node}=[font=\small]
\begin{axis}[
	symbolic x coords = {
		$75$,
		$150$, 
		$225$, 
		$300$, 
	},
	xtick = data,
	%x tick label style={rotate=90, anchor=east},
	xlabel = Edge length of the input patch ($\text{voxel}$),
	y label style={at={(axis description cs:0.08,0.5)},anchor=south},
	ylabel = Mean AUC,
	ymax = 1.0,
	]
	\addplot[ybar, bar width = 20, fill=gray, 
	nodes near coords={\pgfmathprintnumber[fixed zerofill, precision=3]{\pgfplotspointmeta}},]
		coordinates{
		($75$, 0.952)
		($150$, 0.790)
		($225$, 0.722)
		($300$, 0.712)
		
	};	
	\addplot[color=blue, mark=x,]
	coordinates{
		($75$, 0.952)
		($150$, 0.790)
		($225$, 0.722)
		($300$, 0.712)
		
	};	
\end{axis}
\end{tikzpicture}
\caption
{Classification performance for different patch sizes. Enlarging the patch significantly reduces the performance. Smaller patches force the classifier to focus more on the appendix features.}
\label{fig:patchsize}
\end{figure}

As shown in Fig.~\ref{fig:patchsize}, we observed significant decline in the performance for enlarged patches. Because of the tiny structure of the appendix, it is difficult to learn the decision effectively based on the appendix anatomy in larger patches. Smaller patch-size forces the classifier to focus more on the appendix, resulting in greater AUC. This experiment also points to the challenge of using an end-to-end direct image-based diagnosis. The size of the smallest patch used in our experiment was $75 \times 75 \times 75$ voxels. Further shrinking the patch cuts off significant portion of the appendix, therefore, was avoided. We use this smallest patch-size in our final diagnosis model because this gave the best performance. 
\par

\begin{figure*}[t!]
\centering
\includegraphics[scale=1.0]{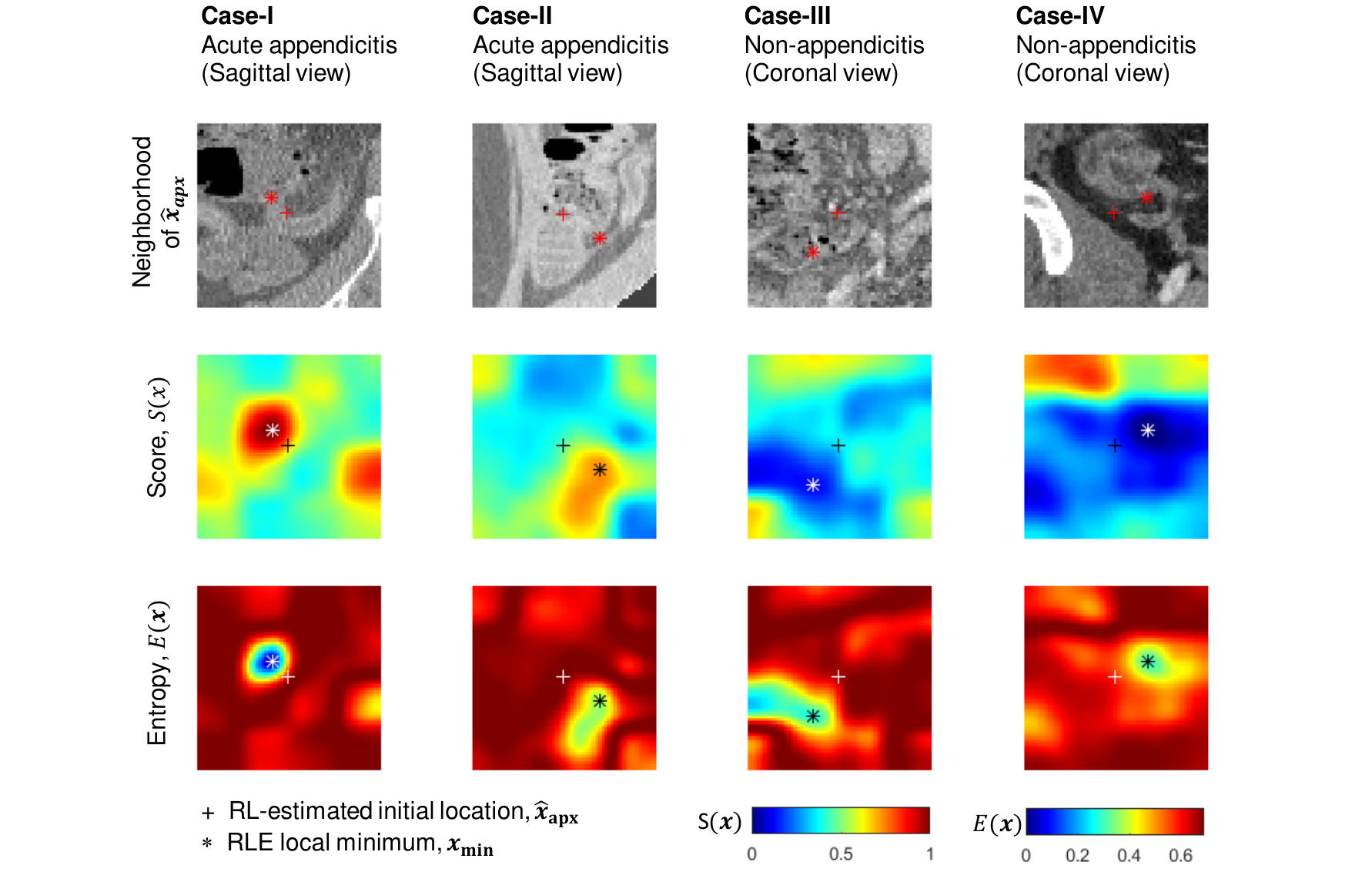}
\caption{Four example cases showing the resultant score and entropy map for the neighborhood of the appendix position localized by the RL-agent. The local minima of the RLEs are also shown along with the indication of the initial appendix localization. Because the appendicitis classifier is trained on the patches centered at the base, the low entropy region is prominent near the base showing a high certainty of the classifier. The score at the initially localized position is not always optimal, whereas, the local minima of the RLE provide the optimal score consistently.}
\label{fig:rle_appx}
\end{figure*}

\subsection{Final Diagnosis Performance}
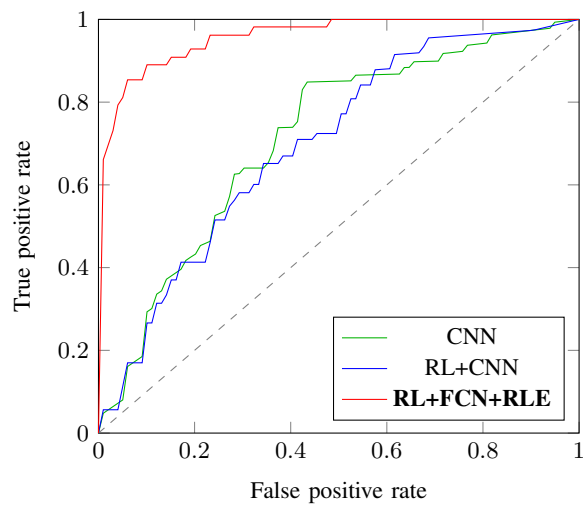
\begin{figure}[h]

\pgfplotsset{
width=0.9\columnwidth,
height=0.8\columnwidth,
}
%\centering
\begin{tikzpicture}
%\pgfplotsset{%
%    width=0.4\textwidth,
%    height=0.4\textwidth,
%}
\tikzstyle{every node}=[font=\small]
\begin{axis}[
    y label style={at={(axis description cs:0.08,.5)},anchor=south},     
    xlabel={False positive rate},
    ylabel={True positive rate}, 
    %xtick distance=600  
    xmin=0, xmax=1,
    ymin=0, ymax=1,
    legend pos=south east,
    %ymajorgrids=true,
   %grid style=dashed,
]
\addplot[mark=none, color=gray, dashed, forget plot]
coordinates{
	(0,0)
	(1,1)
};

\addplot [mark=none,color=ao] table [col sep=comma, mark=none] {auc_cnn.csv};
    \addlegendentry{ CNN}

\addplot [mark=none,color=blue] table [col sep=comma, mark=none] {auc_rl.csv};
    \addlegendentry{ RL+CNN}
    
\addplot [mark=none,color=red] table [col sep=comma, mark=none] {auc_rle.csv};
    \addlegendentry{\bf RL+FCN+RLE}
\end{axis}
\end{tikzpicture}
\caption
{ROC curve comparison. The proposed method (RL+FCN+RLE) shows significant improvement over the RL-only approach (RL+CNN) of directly classifying the RL-localized patch and the single-shot approach (CNN) of directly  classifying the lower-right abdominal sub-volume.}
\label{fig:roc}
\end{figure}

We report the performance of the entire automatic diagnosis framework integrating the initial appendiceal base localization and the RLE-based decision in Table~\ref{tab:final}. For computing the RLE, we obtained the score map for a neighborhood of $120\times 120 \times 70$ voxels centered at the localized base position. This area of the neighborhood is sufficient for detecting the RLE, given the  localization error of about $8$ mm. The resultant score-maps for different test volumes are visualized in Fig.~\ref{fig:rle_appx}. To show the importance of the proposed RLE, we also report the performance of the RL-only diagnosis approach, where the diagnosis is obtained directly based on the patch about the localized position. Furthermore, we include the performance of a standard CNN-only approach for comparison, where an image-based single-shot diagnosis is performed. To give some advantage to this standard approach, we only used the lower-right part of the abdominal CT as the CNN input. Besides presenting the AUC for recognizing the acute appendicitis, we present the optimal sensitivity (i.e., \textit{true positive rate}) and specificity (i.e., \textit{true negative rate}) for each of the methods in Table~\ref{tab:final}. We also plot the receiver-operating characteristics (ROC) curves for all these methods in Fig.~\ref{fig:roc}.

The ROC plots in Fig.~\ref{fig:roc} clearly shows the performance improvement of the proposed method compared with the other methods. The proposed method utilizing the RLE showed a mean AUC as high as $0.961$, which is significantly greater than the AUC of the CNN-only approach. The standard CNN-based approach providing the diagnosis based on the whole lower-right abdomen could only give an AUC of $0.715$. The optimal sensitivity and specificity of the proposed RLE-based method were $0.910$ and $0.926$ respectively, whereas, the CNN-only approach gave a sensitivity of $0.669$ and a specificity of $0.688$. The significant improvement of the proposed method over the standard CNN-based diagnosis can be traced back to the performance dependency on the patch-size discussed earlier. The input used by the end-to-end CNN approach has a tiny region covering the appendix among the comparatively large and complex region of the lower-right abdomen. The proposed RLE-based approach uses a smaller and tighter appendix patch as its core classifier-input, allowing for an easier focus on the appendix reducing the non-appendix portion of the lower-abdomen.
 
On the other hand, the AUC of the RL-only approach was also low despite using the small patch size. With an AUC of $0.698$, the RL-only approach gave an optimal sensitivity of $0.671$ while providing a specificity of $0.647$. The chosen patch-size could provide better performance in the standalone classification test using the manually annotated patch-location. However, obtaining the classification directly from the automatic localization is difficult because of the unavoidable localization error of the RL-agent with respect to the manual annotation. Therefore, using this error-prone automatic patch-location gave a poor diagnosis performance comparatively.

Instead of only using the agent-located patch, the proposed RLE-based decision utilizes neighborhood patches about the localized position of the appendiceal base for obtaining the diagnosis score-map of the local neighborhood. Thus, the proposed method, despite the localization error, can include the score of the desired patch in its score-map. The low-entropy region showed a promising solution to successfully extract this optimal score from the most confident decision-region near the appendiceal base. Fig.~\ref{fig:rle_appx} shows the resultant score and entropy maps about the neighborhood of the localized positions for different test volumes. In this figure, we also mark the localized position and the finally chosen RLE. We observed the RLE close to the appendix with its local minima near the appendiceal base. The core-classifier is trained on the patches centered at the manually annotated base. Therefore, low entropy is observed for the patches having features similar to those base-centered patches. Whereas the diagnosis score at the localized position was not consistently accurate, the proposed RLE could remarkably improve the diagnosis by reaching the optimal position  robustly regardless of the initial localization error. 

The RLE also contributed to the smaller patch-size, which has significant role in improving the classification accuracy. Integrating the localization with classification for having an automated diagnosis is shown to be difficult under the variance of the localized base-position from the manual annotation. Diagnosis becomes highly sensitive to the initial localization. Enlarging the ROI (i.e., patch-size) could be a solution to reduce the sensitivity, trading off the classifier-performance. The proposed RLE showed a fruitful alternative to this by ensuring robust diagnosis under the error-prone localization results, allowing to keep the patch-size smaller.

The widely used sliding window approach to calculate the neighborhood score-map for obtaining the RLE had a high computational cost. The overall computation time (for the score-map and RLE estimation) was about 251 seconds.  The proposed FCN-based score-map estimation for computing the RLE gave significantly more efficient solution requiring only about 4.8 seconds. The total computation of the proposed automatic diagnosis method (including the initial localization) was about 5 seconds.

\setcounter{table}{2}

\begin{table}[]
\caption{Average Computation Time of Different Steps.}

\centering
\renewcommand{\arraystretch}{1.5}
\setlength{\tabcolsep}{6pt}
\begin{tabular}{ l l c}
\hline
 \multirow{2}{*}{ \bf Step } & {\bf Method} & {\bf Avg. Computation Time}\\
& & (Seconds)\\
\hline
\multicolumn{1}{l}{\multirow{1}{*}{ Appendix localization}} & RL & $0.05$ \\
\hline
\multicolumn{1}{l}{\multirow{2}{*}{ RLE computation}} & Sliding window & $251.32$\\
 & {\bf FCN} & $\boldsymbol{4.82}$\\
\hline

\end{tabular}

\label{tab:time}
\end{table}

\section{Conclusion}
\label{sec:conclusion}
We propose an automatic diagnosis method for identifying the acute appendicitis defined as a painful swelling of the vermiform appendix. With no prior approaches for the task in hand, standard image-based diagnosis approaches using CNN also faces challenge because of the tiny structure of the appendix compared with the input abdominal CT volume. To improve the situation, we adopt an RL-based prior localization of the appendix. Training a CNN classifier using small appendix patches, we obtain the diagnosis score for the neighborhood of the localized position. Finally, we introduce the RLE to robustly find the optimal diagnosis score. Experiment with 319 abdominal CT volumes showed that the proposed method showed an improved diagnosis performance resolving the major challenges of the standard diagnosis frameworks when implemented for this task. The key to the improvement was the smaller training patch size to focus more on the appendix reducing non-appendix portion. While suggested prior localization helped reduce the search space, the RLE contributed to robustly reaching the patch to make the optimal decision from the error-prone localization. Thus, the proposed appendicitis diagnosis can be a significant aid to the physicians.

\appendices

\setcounter{table}{0}
\renewcommand{\thetable}{A-\Roman{table}}

\section{Efficient Score Map Computation using FCN}
\label{ap:fcn}
Inspired by the work of Dou et al. \cite{doumicrobleeds}, we adopt the FCN approach to compute the scores for the whole neighborhood of the initial appendix location in a single pass. While the convolution and pooling layers in the traditional CNN can operate on inputs of arbitrary sizes and produces outputs of the corresponding sizes, the fully connected (FC) layers operates on the flattened feature vector of a fixed size and performs simple matrix multiplication to produce an output vector canceling out the spatial dimensions. FCN suggests converting these FC operations into convolution to allow for arbitrary-sized inputs and corresponding spatial outputs.
\par

\begin{table*}[t]
%\begin{adjustwidth}{-1.25in}{0in} % Comment out/remove adjustwidth environment if table fits in text column.
\caption{Patch-based CNN to FCN Conversion. Each Conv and the first FC layer have rectified linear unit (ReLU) activations, whereas the final FC layer is softmax-gated to output the probability of acute appendicitis and non-appendicitis classes. $X$ represents the arbitrary edge length of a cubic input.}
%\centering
\begin{center}
\renewcommand{\arraystretch}{1.5}
\setlength{\tabcolsep}{10pt}
\begin{tabular}{|l c|l c|}
\hline
%\multicolumn{4}{|l|}{\bf Heading1} & \multicolumn{4}{|l|}{\bf Heading2}\\ \hline
\multicolumn{2}{|c|}{\bf Patch-based CNN} & \multicolumn{2}{c|}{\bf FCN} \\
\hline
{\bf Layer} & {\bf Dimension} & {\bf Layer} & {\bf Dimension}\\
\hline
Input & $75\times 75 \times 75 \times 1$ 
	& Input & $X\times X \times X \times 1$\\
Conv $3\times 3 \times 3, 8$ & $75\times 75 \times 75 \times 8$ 
	& Conv $3\times 3 \times 3, 8$ & $X\times X \times X \times 8$ \\
Pool $2\times 2 \times 2, stride:2$ & $37\times 37 \times 37 \times 8$ 
	& Pool $2\times 2 \times 2, stride:2$ & $X/2\times X/2 \times X/2 \times 8$ \\
Conv $3\times 3 \times 3, 8$ & $75\times 37 \times 37 \times 8$ 
	& Conv $3\times 3 \times 3, 8$ & $X/2 \times X/2 \times X/2 \times 8$ \\
Pool $2\times 2 \times 2, stride:2$ & $18\times 18 \times 18 \times 8$ 
	& Pool $2\times 2 \times 2, stride:2$ & $X/4\times X/4 \times X/4 \times 8$ \\
Conv $3\times 3 \times 3, 8$ & $18\times 18 \times 18 \times 8$ 
	& Conv $3\times 3 \times 3, 8$ & $X/4\times X/4 \times X/4\times 8$ \\
Pool $2\times 2 \times 2, stride:2$ & $9\times 9 \times 9 \times 8$ 
	& Pool $2\times 2 \times 2, stride:2$ & $X/8\times X/8 \times X/8 \times 8$ \\
FC $4$ & $4$ 
	& CONV $9\times 9 \times 9, 4$ & $X/8\times X/8 \times X/8 \times 4$ \\
FC $2$ & $2$ 
	& CONV $1\times 1 \times 1, 2$ & $X/8\times X/8 \times X/8 \times 2$ \\

\hline

\end{tabular}
\end{center}
\label{tab:architecture}
%\end{adjustwidth}
\end{table*}

Let us denote the input patch dimension for the traditional 3D CNN by $M\times M\times M\times 1$, considering a cubic patch with a single-channel intensity. Suppose that the output feature volume dimension of the last convolution (or pooling) layer $L_n$ before the FC layer is $M'\times M' \times M' \times P$, assuming cubic kernels with identical stride in all dimensions. Usually, this feature volumes are flattened to a vector of length $N=M'\times M' \times M' \times P$ before feeding it to the FC layer. Thus, the FC layer has $N \times N'$ weights where the output vector is of length $N'$. To obtain the equivalent convolutional layer, we simply reshape the weight matrix to have $M' \times M' \times M' \times P \times N'$ dimension, i.e., $N'$ convolutional kernels of dimension $M' \times M' \times M' \times P$. Employing this convolutional layer directly after $L_n$ without padding results in a spatial output of dimension $1 \times 1 \times 1 \times N'$. The following FC layers can be replaced by using $1 \times 1 \times 1$ convolution. Replacing the FC layers, the converted network can process input of arbitrary shape and produce outputs keeping the spatial dimension. Table~\ref{tab:architecture} shows the traditional 3D CNN and the converted 3D FCN in parallel.
\par

The FCN-generated score map is a coarser version of the original voxel-wise score map of the sliding window approach. Therefore, identifying the RLEs and their local minima in the resultant score map can be problematic. In the approach of Dou et al. \cite{doumicrobleeds}, the actual score-offset in the original input volume is determined by layer-wise backtracking. For any convolution or pooling layer, a spatial position $\boldsymbol{x'}$ in the layer-output can be roughly traced back to the corresponding position $\boldsymbol{x}$ in the layer-input as follows:
\begin{equation}
\boldsymbol{x}  = d \boldsymbol{x'}+\lfloor \frac{k-1}{2} \rfloor 
\end{equation}
where $d$ and $k$ stands for the stride and kernel size. Here, $d$ and $k$ are scalar because they are identical among all spatial dimensions. Usually, no stride is used in convolutional layer (i.e., $s=1$) and padding is also performed to keep the spatial dimension unchanged. Therefore, only pool layer is responsible for altering the spatial dimension, thus requiring sole attention during backtracking. Considering the widely used pool with $s=2$ and $k=2$, the position $\boldsymbol{x}^\text{out}$ in the final FCN can simply be traced back to the approximated original position $\boldsymbol{x}^\text{in}$ as: $\boldsymbol{x}^\text{in} = 2^n\boldsymbol{x}^\text{out}$, where $n$ is total number of pooling layers in the network. Therefore, $2^n$ can also be seen as the required upsampling factor, which we denote by $f$. 
\par
However, scores for only about $\frac{1}{f}$ uniform positions of the input can be found by such backtracking, creating gaps in the resultant score map where no correspondence from the FCN output is found. Considering only one dimension for example, two adjacent positions $x^\text{out}_{i}$ and $x^\text{out}_{i+1}$ in the output space have a difference of $f$ steps in the actual input space, resulting in $2^n-1$ gaps between the backtracked positions $x^\text{in}_{k}$ and $x^\text{in}_{k+f}$.         To fill these gaps, we simply slide the FCN over the input. The process can be depicted by the following scenario for one dimension. For an input vector $\boldsymbol{u} = (u_1, u_2, ..., u_k)$, we collect FCN outputs $\boldsymbol{v}_i$ for slided inputs $\boldsymbol{u}_i = (u_i, u_{i+1}, ..., u_{i+k-f})$, $i\in \mathbb{N}$ and $1\le i \le 2n$. The output with the backtracked indices can be represented by $\boldsymbol{v}_i = (v_i, v_{i+f}, v_{i+2f}, ...)$. Thus, the approximated score map for all the spatial positions can be obtained by fusing the elements of $\boldsymbol{v}_1, \boldsymbol{v}_2, ..., \boldsymbol{v}_{f}$ based on their backtracked index. In order to obtain the upsampled 3D score map, we need to repeat this for all dimensions. Therefore, we need to collect and traverse the FCN for $f^3 = 2^{3n}$ samples by sliding over the original input. For the proposed network with three pooling layers, we should traverse the FCN for 512 samples to fully upsample. However, our goal is only to detect the RLE in the score map rather than to find correspondence for each voxel in the input. For detecting the low-entropy region, upsampling by $6$ times ( instead of the full upsampling factor $f = 8$) give a satisfactory visualization of the entropy map, which only requires 64 samples for FCN.

%\section*{Acknowledgment}
%
%
%This research was supported by Basic Science Research Program through the National Research Foundation of Korea (NRF), funded by the Ministry of Education, Science, Technology (No. 2017R1A2B4004503).

\ifCLASSOPTIONcaptionsoff
  \newpage
\fi



%\bibliographystyle{IEEEtran}

\bibliographystyle{IEEEtran}
\bibliography{mybib}


%\newpage


%\vfill


\end{document}